\documentclass[aps,prl,prabib,twocolumn,showpacs,nofootinbib]{revtex4}
\usepackage{graphicx} \usepackage{amsmath} \usepackage{amssymb}
\usepackage{amsfonts} \usepackage{bm}

\begin{document}

\newcommand{\be}{\begin{equation}} \newcommand{\ee}{\end{equation}}
\newcommand{\bea}{\begin{eqnarray}}\newcommand{\eea}{\end{eqnarray}}

\title {Dipole binding in a cosmic string background due to
quantum anomalies}

\author{Pulak Ranjan Giri} \email{pulakranjan.giri@saha.ac.in}

\affiliation{ Saha Institute of Nuclear Physics, 1/AF Bidhannagar,
Calcutta 700064, India}

\begin{abstract}
We propose quantum dynamics for the dipole moving in cosmic string background
and show that the classical scale symmetry of a particle moving in cosmic
string background
is still restored even in the presence of dipole moment of the particle.
However,
we show that the classical scale symmetry is broken due to inequivalent
quantization of the the non-relativistic system. The consequence of this
quantum anomaly is the formation of bound state in the interval $\xi\in(-1,1)$.
The inequivalent quantization is characterized by a 1-parameter family
of self-adjoint extension parameter $\Sigma$. We show that within the interval
$\xi\in(-1,1)$, cosmic string with zero radius can bind the dipole
 and the dipole does not fall into the singularity.
\end{abstract}



\date{\today}

\maketitle Anomaly  \cite{dono,trei} is a breaking of classical
symmetry  due to quantization of the system, which occurs in various
problems in physics.  It is one of the three possible symmetry
breaking \cite{peskin} i.e., spontaneous, explicit and anomalous
symmetry breaking, which are extremely important due to their
consequences in different physical processes. Chiral anomaly
\cite{bell,adler} is one  such important example in high energy
physics. In quantum field theory  the concept of anomaly has been
used successfully \cite{dono,epele}. The other important area of
physics is  string theory \cite{pol}, where  anomaly has also been
used successfully.  In quantum mechanics, anomaly can be understood
as follows.  An operator, which is the generator of the symmetry in a
classical system,  becomes anomalous when it does not keep the
domain of the Hamiltonian invariant. By this definition of anomaly,
it has been shown that in molecular physics \cite{camblong} (in
quantum mechanical context), there exists interesting scaling
anomaly. For example, interaction of an electron in the field of a
polar molecule is a simple example of anomaly, where the classical
scaling symmetry of the system is broken once it goes inequivalent
quantization \cite{giri}. An obvious consequence of this  scaling
anomaly in molecular physics is the occurrence of bound state and
the the dependence of momentum in the phase shift of the $S$-matrix.

In cosmic string scenario, scaling anomaly has been observed
\cite{giri1} for particles moving in it, where the induced potential
is $1/r^2$ in nature. Inverse square potential appears in various
situations in physics starting from molecular physics to black hole
\cite{camblong,giri,kumar}. The anomaly, shown in  Ref. \cite{giri1}, leads
to bound state for the particle. In this letter we consider quantum
dynamics of a particle with dipole moment $D$ moving in cosmic
string background. This problem has been discussed \cite{lima} for
large negative coupling constant of the inverse square potential,
where the particle  falls into the center \cite{landau} due to the
formation of  infinite number of bound states with ground state
energy being negative infinite. However fall to the center has been
avoided \cite{lima} by considering a finite radius for the cosmic
string. We will however consider a particular portion of the
coupling constant of the inverse square potential which will allow
us to obtain nontrivial boundary condition. This nontrivial boundary
condition will break the scaling symmetry by introducing a length
scale in the form of a single bound state.

Quantum mechanics \cite{cav} in cosmic string background has
received lot of interest due to its analogy \cite{pes} with
Aharanov-Bohm effect \cite{aha}. In relativistic theory it has been
shown \cite{sousa} that Dirac equation in cosmic string background
needs nontrivial boundary condition to be imposed on the spinor
wave-function at the origin. In language of mathematics the
construction of nontrivial boundary condition  is usually  called
self-adjoint extensions \cite{reed}. The extensions can be
characterized by independent parameters and different values of the
parameters lead to inequivalent theories. It has been observed
\cite{wilczek} that in cosmic string scenario the fermionic charge
can be non-integral multiple of Higgs charge. Since the flux is
quantized with respect to the Higgs charge it will lead to
nontrivial Aharanov-Bohm scattering for fermion. The cross section
increases due to this Aharanov-Bohm scattering in addition to
gravitational scattering.  In non-relativistic theory\cite{fil}, the
consideration of inequivalent quantization is also inevitable in
order to get bound state for the particle moving in cosmic string
background. In Ref. \cite{ger1} gravitational scattering by
particles of a spinning source in two dimensions has been studied.
There it  has been shown that the energy eigenvalue and
corresponding eigenfunction of a particle in the field of a massless
spinning source is equivalent to that in a background Aharanov-Bohm
gauge field of an infinitely thin flux tube. This topological defect
appears in astrophysics \cite{pijush} and also in condensed matter
physics \cite{aze}.

This letter has been organized in the following way: First, we study
the scaling symmetry of the classical system, which undergoes
anomalous breaking upon quantization; Second, we make an
inequivalent quantization of the system, which is responsible for
anomaly and discuss its consequences.

First, Scaling symmetry is associated  with the transformation
$\bf{r}\to\lambda\bf{r}$ and $t\to\lambda^2t$, where $\lambda$ is
the scaling factor. In classical physics when the action is
invariant under this transformation, then the corresponding system
has scale symmetry. Since in non-relativistic quantum theory, cosmic
string induces a $V=\frac{(1-\alpha^2)D^2}{48\pi\alpha^2
r^2}\cos2\Theta$ \cite{grats}  potential to the the particle with
dipole moment $D$ moving in its background, the relevant classical
symmetry would be the scale symmetry. To be more specific
classically, the dipole moving in cosmic string background can be
described by the Lagrangian $L=\frac{M}{2}g_{ij}\dot{\bf
r}^{i}\dot{\bf r}^{j} -V$. This Lagrangian $L$ scales as
$\frac{1}{\lambda^2}L$. So the action $\mathcal A= \int dt L$ will
be scale invariant under the transformation $\bf{r}\to\lambda\bf{r}$
and $t\to\lambda^2t$. The scale invariance of this action means, if
$\psi$ is an eigenstate of the Hamiltonian
$H=\frac{M}{2}g_{ij}\dot{\bf r}^{i}\dot{\bf r}^{j} +V$  with
eigenvalue $E$, i.e., $H\psi= E\psi$, then
$\psi_\lambda=\psi(\lambda \bf r)$ will also be an eigenstate of the
same Hamiltonian with energy $E/\lambda^2$. This essentially means
that the system with scale symmetry does not have any lower bound in
energy and therefore cannot have any bound state. Scale symmetry is
however a part of larger conformal symmetry formed  by three
generators: the Hamiltonian $H$, the Dilatation generator $\mathcal
D= tH- \frac{1}{4}(\bf{r}.\bf{p}+ \bf{p}.\bf{r})$ and the conformal
generator $K= Ht^2-\frac{1}{2}(\bf{r}.\bf{p}+ \bf{p}.\bf{r}) +
$$\frac{1}{2}M\bf{r}^2$. They form the $SO(2,1)$ algebra:
$[\mathcal D,H]= -i\hbar H$, $[\mathcal D,K]= i\hbar K$, $[H,K]=
2i\hbar \mathcal D$ \cite{wyb}. We will show in our case that this
scale symmetry will break once the the classical system is
quantized.
\begin{figure}
\includegraphics[width=0.45\textwidth, height=0.15\textheight]{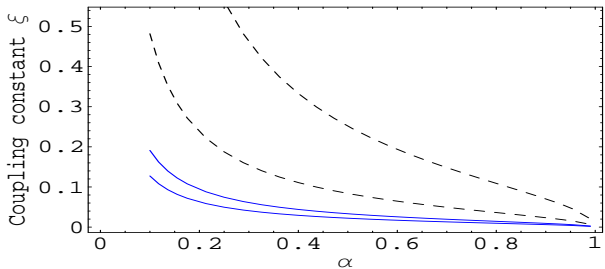}
\caption {(color online) A plot of the coupling constant $\xi$ as a
function of the cosmic string parameter $\alpha$. Dotted graphs
correspond to $\Theta=\pi/8$ and from top to bottom $D=0.5, 1.5$
respectively. Solid  graphs (blue) correspond to $\Theta=\pi/5$ and 
from top to bottom $D=0.2, 0.3$ respectively.}
\end{figure}
\begin{figure}
\includegraphics[width=0.45\textwidth, height=0.15\textheight]{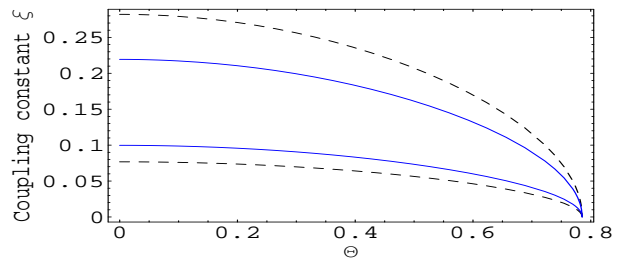}
\caption {(color online) A plot of the coupling constant $\xi$ as a
function of
  $\Theta$. Dotted graphs correspond to $D=0.5$ and 
from top to bottom
  $\alpha=0.2,0.6$ respectively. Solid graphs (blue) correspond to 
$\alpha=0.5$ and from top to
  bottom $D=0.5, 1.1$ respectively.}
\end{figure}

We consider a non-relativistic particle of mass $M$, dipole moment $D$
moving in the
background field of cosmic string. The background is described by
the space-time metric in cylindrical coordinate $(r,\phi,z)$ as
\begin{eqnarray}
ds^2 = dt^2- dz^2- dr^2 - \alpha^2r^2d\phi^2\,,
\end{eqnarray}
where $\alpha=1- 4G\mu <1 $ characterizes the string, with $\mu$ is
the mass per unit length of the string. The constant $\alpha$
introduces an angular deficit of $2\pi(1-\alpha)$ in the Minkowski
space-time. The interaction between the dipole and the cosmic string
background is described by the electromagnetic self-energy
\cite{grats} of the dipole due to the non-flat geometry. The
potential induced in the non-relativistic system is
$V=\frac{(1-\alpha^2)D^2}{48\pi\alpha^2
r^2}\cos2\Theta$ \cite{grats}, where $\Theta$ is the angle between
the string and the dipole moment. This potential transforms under the
scale transformation $\bf r= \lambda\bf r$ and $t=\lambda^2 t$
in such a way that the Schr\"{o}dinger equation for the system becomes
scale covariant. Due to cylindrical symmetry of
the space, we can easily see that the motion of the particle in the
z direction is basically a free particle motion, described by the
wave-function $e^{ikz}$, where $k$ is wave-vector of the particle along
the $z$ direction. Since we are considering an infinite cosmic
string along the $z$ direction, it is enough to discuss the motion
of the particle on the plane perpendicular to the $z$ direction. The
motion of the particle on the plane perpendicular to the $z$ axis is
described by the time independent Schr\"{o}dinger equation (in
$\hbar^2=M=1$ unit)
\begin{eqnarray}
\left(-\frac{1}{2}\nabla^2
+\frac{(1-\alpha^2)D^2}{48\pi\alpha^2r^2}\cos2\Theta\right)\Psi=
E\Psi \label{schrodinger}
\end{eqnarray}
The wave-function can be separated as $\Psi(r,\phi) =
R(r)\exp(im\phi)$ and (\ref{schrodinger}) gives the radial
equation
\begin{eqnarray}
 H_D R(r) \equiv -\left ( \frac{d^2}{dr^2} + \frac{1}{r}\frac{d}{dr}
-\frac{\xi^2}{r^2} \right )R(r) = 2E R(r), \label{radial}
\end{eqnarray}
where $H_D$ is the radial Hamiltonian, with $\xi^2 =
\frac{1}{\alpha^2}\left
(\frac{(1-\alpha^2)D^2}{24\pi}\cos2\Theta-m^2\right)$ and $m=0,\pm
1, \pm 2, \cdots$. We will now discuss the solution of the
Hamiltonian $H_D$.

To discuss that we need to know some general properties of an
operator, let say $A$. In this article, let us restrict ourselves to
the case of unbounded operator, because the Hamiltonian we are
discussing is unbounded. Now,  for an unbounded operator $A$, one
can define a domain $D(A)$, such that the domain is dense in the
Hilbert space.  From the information of $A$ and $D(A)$, one can
construct the adjoint operator $A^*$ and the corresponding domain
$D(A^*)$ by using the relation
$\int_0^\infty\phi^*(r)A\chi(r)dr=\int_0^\infty\left(A^*\phi(r)\right)^*\chi(r)dr$,
$\forall \chi(r)\in D(A)$. The condition for self-adjointness for
the operator $A$ is given by $D(A)= D(A^*)$. An alternative
definition of self-adjointness is given in terms of deficiency
indices, found by using von Neumann's method.  According to von
Neumann's method, the deficiency indices $n_\pm$ are defined by
dimension of the kernel $Ker(i\pm A^*)$. If $n_\pm=0$, then the
operator $A$ is essentially self-adjoint. If $n_+=n_-=n\neq 0$, then
$A$ is not self-adjoint but admits self-adjoint extensions.
Self-adjoint extensions can be characterized by $n^2$ parameters.
Different values of the parameters give rise to different physics.
For, $n_+\neq n_-$, the operator $A$ cannot have any
self-adjoint extensions.
\begin{figure}
\includegraphics[width=0.45\textwidth, height=0.15\textheight]{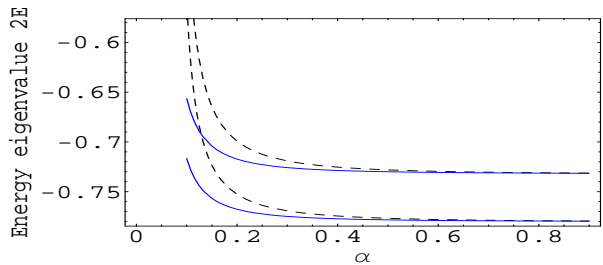}
\caption {(color online) A plot of the energy eigenvalue $2E$  as a
function of the cosmic string parameter $\alpha$. Dotted graphs
correspond to $\Theta=\pi/8$, $D=1.6$ and from top to bottom
$\Sigma=\pi/8, \pi/10$ respectively. Solid graphs (blue) correspond to
$\Theta=\pi/12$, $D=1$
 and from top to bottom $\Sigma=\pi/8,
\pi/10$ respectively.}
\end{figure}

Let us now come back to the discussion  our operator of interest,
which is $H_D$. The Hamiltonian $H_D$ acts over the Hilbert space of
square-integrable functions, given by the domain $\mathcal{L}^2[R^+,
rdr]$. As discussed earlier, classically this system is scale
invariant under the scale transformation $\bf r=\lambda r$,
$t=\lambda^2 t$. It can also be understood from the fact that the
coupling constant of the inverse square potential $\xi$ is
dimensionless coefficient. Now, we need to see whether this scale
symmetry is still restored after quantization. Since this kind of
model (inverse square interaction) has been investigated extensively
in literature, we know that the Hamiltonian $H_D$ is essentially
self-adjoint for $\xi^2\geq1$. Since any system is defined by a
Hamiltonian and its corresponding domain, in our case the
Hamiltonian $H_D$ for  $\xi^2\geq1$ acts over the  domain
\begin{eqnarray}
\mathcal{D}_0=\{\psi\in \mathcal{L}^2(rdr), \psi(0)=\psi'(0)=0\}
\end{eqnarray}
Let us now investigate another portion of the coupling constant
$\xi^2$. In this article we will not consider the strong region,
because it has been investigated earlier. So the remaining region
left to be investigated is $\xi\in(-1,1)$. In this region the
Hamiltonian is not essentially self-adjoint and therefore we need to
make self-adjoint extensions of the original domain, so that the
Hamiltonian becomes self-adjoint. For the moment we consider
$\xi\neq 0$, because the case $\xi=0$  should be treated separately.
From now onward we confine our analysis to the zero angular momentum
states, i.e., we set $m=0$ in the expression of $\xi$ for
simplicity. However our analysis is valid as long as $0\leq
\xi^2<1$.  Since the quantum dynamics in the interval $\xi\in(-1,1)$
essentially depends on the behavior of the coupling constant $\xi$,
we  plot $\xi$ as a function of the cosmic string parameter $\alpha$
in FIG. 1 and in FIG. 2, we plot the same thing as a function of the
variable $\Theta$. Note that for $\xi\in(-1,1)$, the deficiency
indices are $(1,1)$, so the self-adjoint extensions are
characterized by a 1-parameter element $e^{i\Sigma}$. Now, the
domain under which our Hamiltonian $H_D$ should be self-adjoint  is
given by $\mathcal {D}_\Sigma$. Mathematically this domain can be
represented by
\begin{eqnarray}
\mathcal {D}_\Sigma = \{\mathcal {D}_0 +\phi_+ + e^{i\Sigma}
\phi_-\}\,,
\label{d6}
\end{eqnarray}
where the deficiency space
solutions $\phi_\pm$ are
\begin{eqnarray}
\phi_+ = K_\xi(re^{-i\pi/4})\,,~~ \phi_- = K_\xi(re^{+i\pi/4})\,,
\end{eqnarray}
where $K_\xi$ is the modified Bessel function \cite{abr}. The
behavior of any function, belonging to the domain $\mathcal {D}_\Sigma$,
near singularity  $r\to 0$ can be found from the
behavior  of $\phi_+ + e^{i\Sigma}\phi_-$ at short distance, because
near singularity, functions belonging to the domain $\mathcal {D}_0$ goes to
zero. Therefore
\begin{eqnarray}
\phi_+ + e^{i\Sigma}\phi_-\simeq
\mathcal{A}_+\left(\frac{r}{2}\right)^\xi +
\mathcal{A}_-\left(\frac{2}{r}\right)^{\xi}
\end{eqnarray}
where, $\mathcal{A}_\pm= -\frac{\pi i}{\sin(\pi\xi)}
\frac{\cos(\frac{\Sigma}{2}\pm \frac{\pi\xi}{4})}{\Gamma(1 \pm\xi)}$.

Let us now solve the eigenvalue problem (\ref{radial}). Since  for
$\xi^2\geq 1$, the Schr\"{o}dinger equation does not have any
normalizable solutions, we can't have any bound state, because bound
state solutions must be normalized in quantum mechanics. On the
other hand it can be shown that \cite{kumar} for $\xi\in(-1,1)$,
there is exactly one bound state with energy $2E$,  and
eigenfunction $R(r)$:
\begin{eqnarray}
2E= -\sqrt[\xi]{\frac{\cos\frac{1}{4}\left(2\Sigma+ \xi\pi\right)}
{\cos\frac{1}{4}\left(2\Sigma- \xi\pi\right)}}\,,~~ R(r)=
K_\xi(\sqrt{2E} r)
 \label{bound1}
\end{eqnarray}
The bound state energy $2E$ in (\ref{bound1}) as a function of the cosmic
string
parameter $\alpha$ has been plotted in FIG. 3. In FIG. 4, the
eigenvalue $2E$ has been plotted as a function of the dipole moment
$D$.

As pointed out before, according to scale symmetry there should
not have any bound state solution. But in our system we get bound state
solution for nontrivial boundary condition. The bound state
eigenvalue can be considered as a scale in the system, which has
emerged due to nontrivial boundary condition. Thus the classical
scale symmetry is destroyed after quantization of the system. Let us
now discuss the scaling anomaly in terms of operators over the
Hilbert space. In quantum mechanics there is a operator called
scaling operator, which encodes the features of scaling symmetry.
The scaling operator is given by
\begin{eqnarray}
\Lambda= \frac{1}{2}(rp + pr)\,,~~\mbox{where}~p= -i\frac{d}{dr}
\end{eqnarray}
One can check that this scaling operator $\Lambda$ is symmetric on the domain
$\mathcal {D}_0$ of the Hamiltonian $H_D$. It can also be checked that
for $\xi^2\geq 1$, the domain of the Hamiltonian $H_D$ remains invariant when 
$\Lambda$ when acts on it. For
$\xi^2\in(-1,1)$, $\Lambda\phi= -\frac{i}{2}\left(\phi +
2r\phi'\right)$,
where $\phi$ is any element, belonging to the domain
$\mathcal{D}_\Sigma$. The behavior of the function
$\Lambda\phi$ near singularity ($r\to 0$)  can be found as
\begin{eqnarray}
\Lambda\phi \simeq -\frac{i}{2}
\left[(1+2\xi)\mathcal{A}_+\left(\frac{r}{2}\right)^\xi  +
(1-2\xi)\mathcal{A}_-\left(\frac{2}{r}\right)^{\xi}\right]\label{anomaly}
\end{eqnarray}
where, the constants $\mathcal{A}_\pm$ are defined above.

Comparing the expression (\ref{d6}) and (\ref{anomaly}), we see that
$\Lambda\phi$  does not leave the domain of the
Hamiltonian invariant, due to the two different terms $(1+2\xi)$ and
$(1+2\xi)$ in the expression (\ref{anomaly}).
Scaling symmetry is  thus broken  anomalously. The reason for this anomaly is
the inequivalent quantization of the system, by making a self-adjoint
extensions of the initially non self-adjoint Hamiltonian. Note that not all
values of the self-adjoint extensions parameter $\Sigma$ give rise to scaling
anomaly. There are some values of the parameters for which scale symmetry is
restored even after quantization.
For example, for $\Sigma=
(1\pm\frac{\xi}{2})\pi$ there is no  bound state. One can also check
from (\ref{anomaly}) that for $\Sigma=
(1\pm\frac{\xi}{2})\pi$,   $\Lambda$ leaves the
domain of the Hamiltonian invariant.
\begin{figure}
\includegraphics[width=0.45\textwidth, height=0.15\textheight]{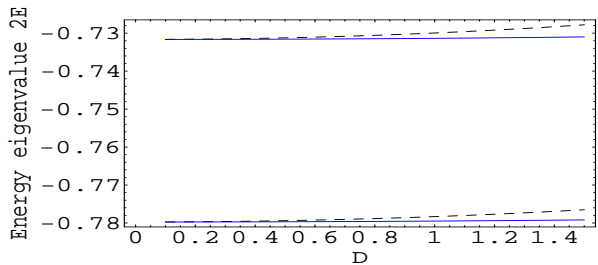}
\caption{(color online) A plot of the energy eigenvalue $2E$  as a
function of the dipole moment $D$. Solid  graphs (blue) 
correspond to
$\Theta=\pi/10$, $\alpha=0.8$ and from top to bottom $\Sigma=\pi/8,
\pi/10$ respectively. Dotted graphs correspond to 
$\Theta=\pi/12$,
$\alpha=0.5$
 and from top to bottom $\Sigma=\pi/8,
\pi/10$ respectively.}
\end{figure}

The case for $\xi = 0$  can be handled in a
similar fashion. The bound state energy and the wave function in this case are
given by \be2 E = - {\exp}\left[\frac{\pi}{2}  {\cot}
\frac{\Sigma}{2}\right]\,,~~ R(r) =
K_0\left( \sqrt{-2E} r\right)\,. \ee respectively, where $K_0$ \cite{abr}
is the
modified Bessel function. Here also the existence of bound state
imply breaking of scale symmetry.

In conclusion, we have shown that the presence of dipole moment of a
particle, moving in cosmic string background,  does not break the
classical scale symmetry, which was present without the dipole
moment. However, scale symmetry is anomalously broken by the
inequivalent quantization of the system.  The inequivalent
quantization is characterized by  one parameter family of
self-adjoint extensions. The consequence of this anomaly is the
existence of bound state for the dipole and the scale is provided by
the bound state energy. We have shown that scale symmetry can be
restored for $\Sigma= (1\pm\frac{\xi}{2})\pi$ even after
quantization.

\end{document}